\documentclass[preprint,showpacs,preprintnumbers,amsmath,amssymb,aps,pre]{revtex4-1}
\usepackage{graphicx}
\usepackage{dcolumn}
\usepackage{bm}

\begin{document}

\title{Skewness and kurtosis analysis for non-Gaussian distributions}

\author{Ahmet Celikoglu$^{1,}$}
 \email{ahmet.celikoglu@ege.edu.tr}
\author{Ugur Tirnakli$^{1,2,}$}
 \email{ugur.tirnakli@ege.edu.tr}

\affiliation{
$^1$Department of Physics, Faculty of Science, Ege University, 35100 Izmir, Turkey\\
$^2$Division of Statistical Mechanics and Complexity,
Institute of Theoretical and Applied Physics (ITAP) Kaygiseki Mevkii,
48740 Turunc, Mugla, Turkey\\}

\date{\today}

\begin{abstract}
In a recent paper [\textit{M. Cristelli, A. Zaccaria and L. Pietronero, Phys. Rev. E 85, 066108 (2012)}], 
Cristelli \textit{et al.} analysed relation between skewness and kurtosis for complex dynamical systems 
and identified two power-law regimes of non-Gaussianity, one of which scales with an exponent of 2 and the other 
is with $4/3$. Finally the authors concluded that the observed relation is a universal fact in complex dynamical systems. 
Here, we test the proposed universal relation between skewness and kurtosis with large number of synthetic data 
and show that in fact it is not universal and originates only due to the small number of data points in the data 
sets considered.  
The proposed relation is tested using two different non-Gaussian distributions, namely $q$-Gaussian and Levy distributions. 
We clearly show that this relation disappears for sufficiently large data sets provided that the second 
moment of the distribution is finite. 
We find that, contrary to the claims of Cristelli \textit{et al.} regarding a power-law scaling regime, 
kurtosis saturates to a single value, which is of course different 
from the Gaussian case ($K=3$), as the number of data is increased. On the other hand, if the second moment of the 
distribution is infinite, then the kurtosis seems to never converge to a single value. 
The converged kurtosis value for the finite second moment distributions and the number of data points needed to 
reach this value depend on the deviation of the original distribution from the Gaussian case. 
We also argue that the use of kurtosis to compare distributions to decide which one deviates from the Gaussian more 
can lead to incorrect results even for finite second moment distributions for small data sets, whereas it is 
totally misleading for infinite second moment distributions where the difference depends on $N$ for all finite $N$.   
\end{abstract}


\maketitle


\section{Introduction}

\hspace{0.5cm} 
In the last decades of the 19th century W.F.R. Weldon had encountered a problem while he was analysing the 
evolution of sea shells and collected morphological data. The distribution of this data set was not in the 
shape of a Gaussian. 
This situation was unusual for those days and has brought to mind an important question: Does this deviation 
from the Gaussian stem from a data collection error or the evolution is really going on such a way? 
When the problem was taken up by Pearson, the story which led to the introduction of kurtosis index began. 
For further information, one can see \cite{Zenga} and references therein. 
After firstly introduced by Pearson in 1905 \cite{Pearson}, kurtosis has became a quantity widely encountered 
in many textbooks. Nowadays there is a perception that having a larger kurtosis value means a larger deviation 
from Gaussian distribution. Two non-Gaussian distributions are compared to each other based on this perception. 
There are some common misconceptions like this about using kurtosis. DeCarlo has addressed some of these 
misconceptions and their explanations \cite{DeCarlo}. Various misconceptions and misunderstandings are discussed 
with examples not only from undergraduate level textbooks but also from graduate level ones.

In a recent work, the relation between skewness and kurtosis has been studied by 
Cristelli \textit{et al.} \cite{Cristelli}. 
The authors have analysed three different non-Gaussian data sets. Two of them are taken from The Global 
Centroid-Moment-Tensor (CMT) and ISIDe (Italian catalog) earthquake catalogs. 
For the third data set, they have focused on the daily price returns from the S$\&$P $500$ index. The procedure 
used to analyse the data sets is to divide datasets into subsamples and calculate skewness ($S$) and kurtosis ($K$) 
for each subsample window using the standard definitions of these quantities given by 

\begin{equation}
S=\dfrac{1}{\sigma^3} \left[\dfrac{1}{N} \sum_{i=1}^{N}(x_{i}-\mu)^{3}\right],
\label{skewness}
\end{equation}

\noindent and 

\begin{equation}
K=\dfrac{1}{\sigma^4} \left[\dfrac{1}{N} \sum_{i=1}^{N}(x_{i}-\mu)^{4}\right],
\label{kurtosis}
\end{equation}

\noindent where $N$ is the number of data points, $\mu$ is the mean of the sample and $\sigma$ is the standard deviation. 
The largest dataset (financial data) is divided into subsamples of length $N=250$ and as clearly seen in Fig.~2 
of \cite{Cristelli}, two different scaling regimes of power-law type are observed for kurtosis versus skewness plot. 
In one of these regimes, all points are clustered like a power-law with an exponent of 2, namely, 

\begin{equation}
K=S^{2}+\dfrac{189}{125}
\label{parabol}
\end{equation}
around the point $S=0$ and $K=3$ which are characteristic values of an infinite Gaussian distribution. 
The constant term in Eq.~(\ref{parabol}) is the lower bound for the difference $K-S^{2}$. Although the shape of 
the distribution affects the value of the bound, Pearson has found that it is approximately $1$. In $2000$, 
Klaassen put this relation into its final form as $(K-S^{2}\geq189/125)$, which is in Eq.~(\ref{parabol}), 
for unimodal distributions \cite{Klaassen}.

Outside this regime, the relation between skewness and kurtosis turns out to be a power-law with an exponent of $4/3$. 
The argument given in \cite{Cristelli} in order to explain this behaviour is the following. 
If there is a sufficiently extreme event in the data set, this event dominates the summation and the contribution 
of other points can be considered as negligible. Therefore, moments are given approximately as,

\begin{equation}
S\simeq \frac{\frac{1}{N}(x-\mu)^{3}}{\sigma^{3}} ,
\label{yaklasımS}
\end{equation}

\begin{equation}
K\simeq \frac{\frac{1}{N}(x-\mu)^{4}}{\sigma^{4}} ,
\label{yaklasımK}
\end{equation}


\noindent where $x$ is the value of extreme event. 
One can easily find from Eq.~(\ref{yaklasımS}) that  $(x-\mu)/\sigma\simeq(NS)^{1/3}$ 
and using this expression in Eq.~(\ref{yaklasımK}), the power law relation can be obtained as  

\begin{equation}
K \simeq N^{1/3}S^{4/3} .
\label{power}
\end{equation}
The value of $N$ is 100 for earthquakes and 250 for financial time series \cite{Cristelli}. At this point, the crucial 
question that should be asked is whether the behaviour remains the same as $N$ increases.

Our main purpose here is to test the relation between skewness and kurtosis proposed as being universal 
in \cite{Cristelli} using very large synthetic data sets which are known to be non-Gaussian. 
The other purpose is to find an answer to the following questions: 
(i)~ does larger kurtosis value between any two non-Gaussian distribution always mean larger deviation from Gaussian ? 
(ii)~if not, then how can we compare two different non-Gaussian distributions and decide which one has larger 
deviation from Gaussian ?

\section{$q$-Gaussians as non-Gaussian distributions}

\subsection{Generating $q$-Gaussian Distributions}

There are different methods in the literature to generate Gaussian distributions. One of the most popular and 
well known one is the Box-Muller Method \cite{Box-Muller}. On the other hand, there are many complex systems in nature 
which do not exhibit Gaussian distributions. 
In the literature, there are several examples of experimental, observational and model systems in physics, biology, 
geophysics, economics etc which exhibit $q$-Gaussian distributions. These distributions optimize the nonadditive 
entropy $S_q$, which is defined to be $S_{q} \equiv \left(1- \sum_i p_i^q\right)/ \left(q-1\right)$ and are known 
to be the basis of nonextensive statistical mechanics \cite{Tsallis88,tsallisbook} and recovers the standard 
Boltzmann-Gibbs entropy as a special case when $q\rightarrow 1$.

If $1<q<3$, $q$-Gaussian distributions are long tailed non-Gaussian distributions similar to those as observed for daily 
price returns of economics \cite{TsallisEcomony,Polish} as well as return distributions of 
earthquakes \cite{caruso,ahmet}. Therefore, they are very good candidates to use in order for achieving our purposes 
explained above. These distributions are known to have finite (infinite) second moments for $1<q<5/3$ ($5/3<q<3$). 
Needless to say, one will need a generalization of the Box-Muller method from where $q$-Gaussian 
distributions can be generated. This generalization has been performed by Thistleton \textit{et al.} 
in 2007 \cite{qBox-Muller}. 
Suppose that $U_{1}$ and $U_{2}$ are independent random variables chosen from uniform distribution defined on $(0,1)$. 
It is shown that two random variables $Z_{1}$ and $Z_{2}$ can be defined as

\begin{eqnarray}
Z_{1}&\equiv &\sqrt{-2 \ln_{q^{\prime}} (U_{1})} \cos(2\pi U_{2})\nonumber\\
Z_{2}&\equiv &\sqrt{-2 \ln_{q^{\prime}} (U_{1})} \sin(2\pi U_{2})
\label{q gaussian variables}
\end{eqnarray}
and each of them is a standard $q$-Gaussian deviate characterized by a new parameter $q$, which is given by 
$q=\dfrac{3q^{\prime}-1}{q^{\prime}+1}$. 
Here $\ln_{q}$ is the $q$-logarithm and is defined as
\begin{equation}
\ln_{q}(x)\equiv\dfrac{x^{1-q}-1}{1-q} \qquad x>0 ,
\label{lnq}
\end{equation}
whose inverse is known as the $q$-exponential and is given as

\begin{equation}
\label{qexp} 
e_{q}^{x} \equiv \left\{\begin{array} {c} [1+(1-q)x]^{\frac{1}{1-q}}, \qquad1+(1-q)x\geq0, \\ 
0,\qquad\qquad\qquad else.  \end{array}\right. 
\end{equation} \noindent

\noindent
Finally, one can define $q$-Gaussian distribution as
\begin{equation}
p(x;\mu_{q},\sigma_{q})=A_{q}\sqrt{B_{q}} [1+(q-1)B_{q}(x-\mu_{q})^{2}]^{\frac{1}{1-q}},
\end{equation}  
where $\mu_{q}$ is the $q$-mean value, $\sigma_{q}$ is the $q$-variance, $A_{q}$ is the normalization factor 
and $B_{q}$ is a parameter which characterizes the width of the distribution. These parameters are defined as follows: 
\begin{equation}
\mu_{q} \equiv \dfrac{\int x[p(x)]^{q}dx}{\int[p(x)]^{q}dx}
\label{mq}
\end{equation}

\begin{equation}
\sigma^{2}_{q} \equiv \dfrac{\int (x-\mu_{q})^{2}[p(x)]^{q}dx}{\int[p(x)]^{q}dx}
\label{sigmaq}
\end{equation}

\begin{equation}
\label{Aq} 
A_{q}= \left\{\begin{array} {c} \frac{\Gamma\left[\frac{5-3q}{2(1-q)}\right]}
{\Gamma\left[\frac{2-q}{1-q}\right]}\sqrt{\frac{1-q}{\pi}}, \qquad q<1,  \\ 
\frac{1}{\sqrt{\pi}}, \qquad\qquad\qquad q=1, \\

\frac{\Gamma\left[\frac{1}{q-1}\right]}{\Gamma\left[\frac{3-q}{2(q-1)}\right]}
\sqrt{\frac{q-1}{\pi}},\qquad 1<q<3. \end{array}\right. 
\end{equation} \noindent

\begin{equation}
B_{q}=[(3-q)\sigma^{2}_{q}]^{-1}  \qquad q\in (-\infty,3).
\end{equation}
Using this generalized Box-Muller method, one can generate arbitrarily large number of data sets for 
$q$-Gaussian distributions with any $q$ value. 

\begin{figure*}[t]
\begin{center}
\includegraphics[width=10cm]{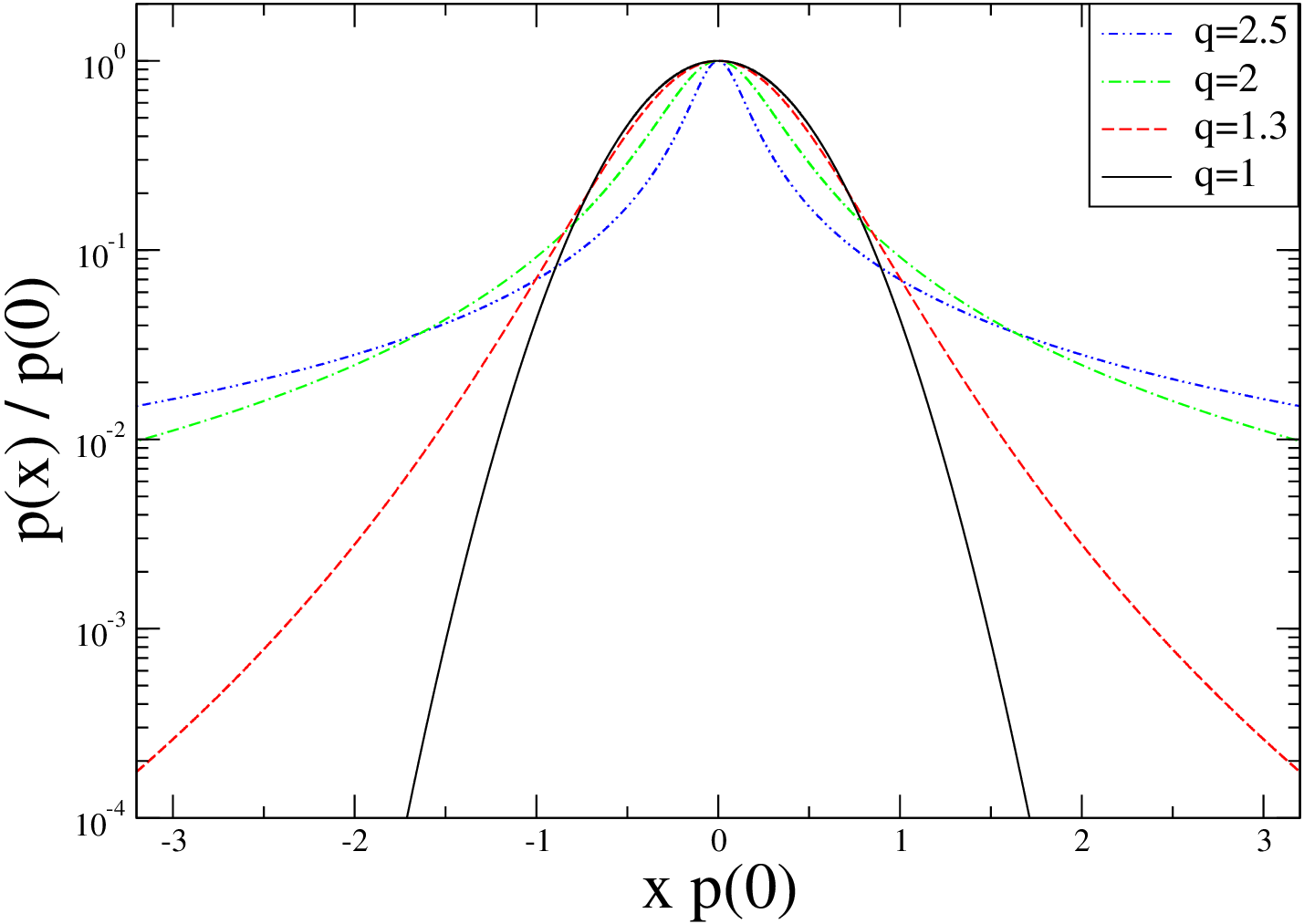}
\end{center}
\caption{(Color online) $q$-Gaussian distributions for representative $q$ values.  
Gaussian distribution is also given as a special case with $q=1$.}
\label{fig:fig1}
\end{figure*}

\subsection{Skewness and Kurtosis}

\begin{figure*}[t]
\begin{center}
\includegraphics[width=10cm]{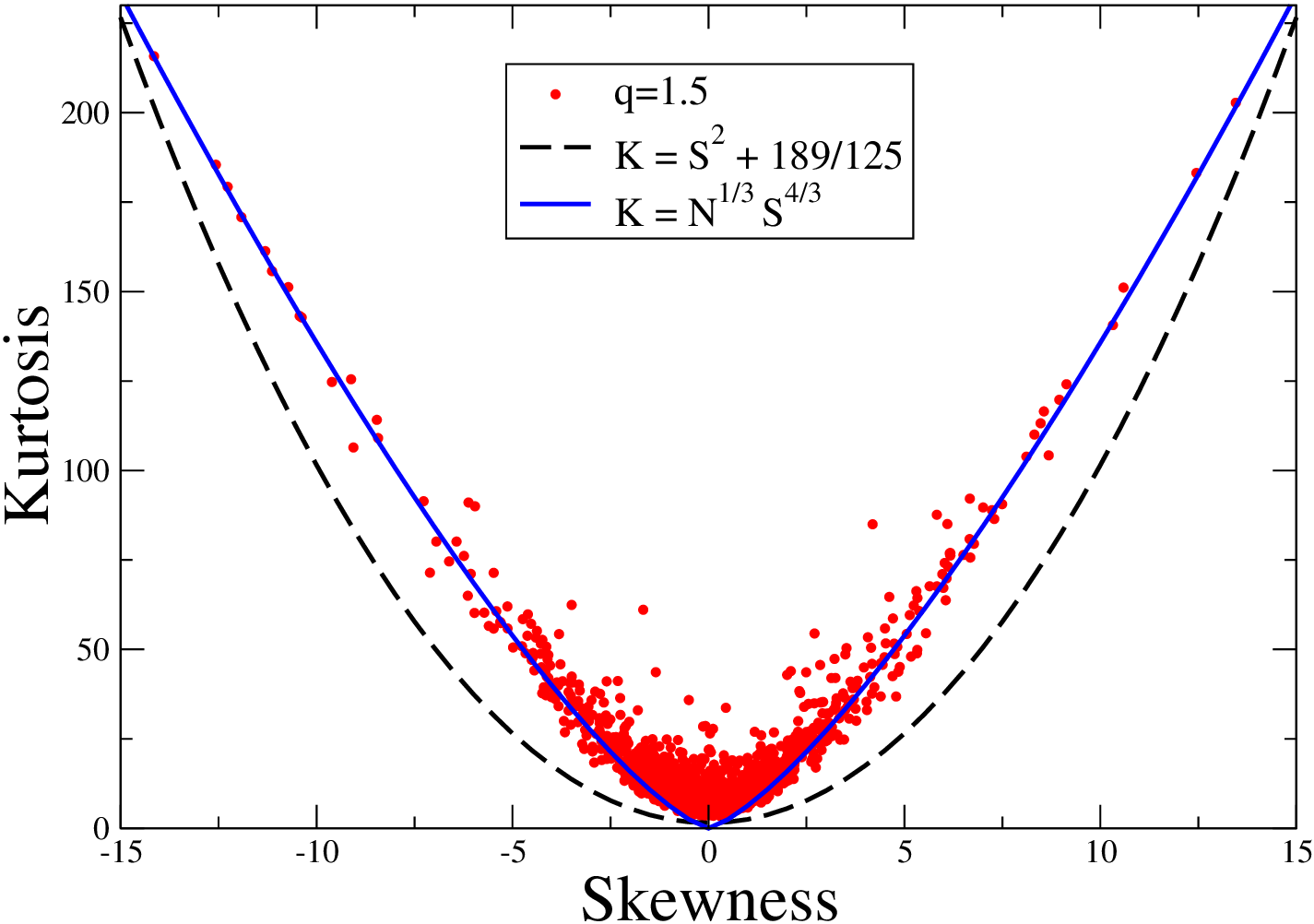}
\end{center}
\caption{(Color online) Kurtosis versus skewness for $q$-Gaussian distribution with $q=1.5$. 
The number of points is $2000$. 
Each of them refers to a sample window and each window has $N=250$ data points. This figure should be compared 
to the Fig.2 of \cite{Cristelli}.}
\label{fig:fig2}
\end{figure*}

The standard definitions of skewness and kurtosis have already been given in Eq.~(\ref{skewness}) and 
Eq.~(\ref{kurtosis}) respectively. Changing the value of $q$ in Eq.(\ref{q gaussian variables}), one can 
simply generate $q$-Gaussian distributions for different $q$ values as seen in Fig.~\ref{fig:fig1}. 
Therefore, now we have all the necessary ingredients to test the proposed relation between skewness and kurtosis. 
We have generated $q$-Gaussian distributions for various $q$ values and divided the datasets to subsamples and calculated 
skewness and kurtosis values for each subsample window, which is exactly the same procedure used in \cite{Cristelli}. 
In order to mimic exactly the results given in Fig.~2 of \cite{Cristelli}, we take $N=250$ for each window and plot 
kurtosis as a function of skewness for $q$-Gaussian distribution with $q=1.5$ in Fig.~\ref{fig:fig2}, where each point 
refers to a window. If Fig.~2 of \cite{Cristelli} and our Fig.~\ref{fig:fig2} are compared, one can easily see that 
they are almost the same. If a zoom is made to the region very close to $S=0$, it is seen that, for this very narrow 
region, data points obey a 
power-law relation with exponent 2. In this regime, it seems that the distributions obtained from each window are not 
very far away from Gaussian. In fact, even if we generate $q$-Gaussian distribution, since the number of data points is 
very small, the tails of the distribution are poorly sampled whereas its central part is sampled highly. 
Since any $q$-Gaussian will not differ very much from Gaussian at its central part, this explains why we see such 
a regime where the points are clustered close to $S=0$ and $K=3$ in $S-K$ plane. Since we generate symmetric 
$q$-Gaussian distributions, one can easily find points with zero skewness, but kurtosis values are greater than 
$3$ for all points in Fig.~\ref{fig:fig2} due to non-Gaussianity. 
When extreme events happen to appear (i.e., data comes from the tails) and become dominant in any window, the data point 
corresponding to this window in $S-K$ plane starts to move away from the $S=0$ and $K=3$ regime. 
As we explained in the Introduction (see Eqs.~(\ref{yaklasımS}) - (\ref{power})), 
if there are sufficiently many extreme events, they dominate all the summation and the relation turns out to be another 
power-law with exponent $4/3$.

\begin{figure*}[!th]
\begin{center}
\includegraphics[width=8.1cm]{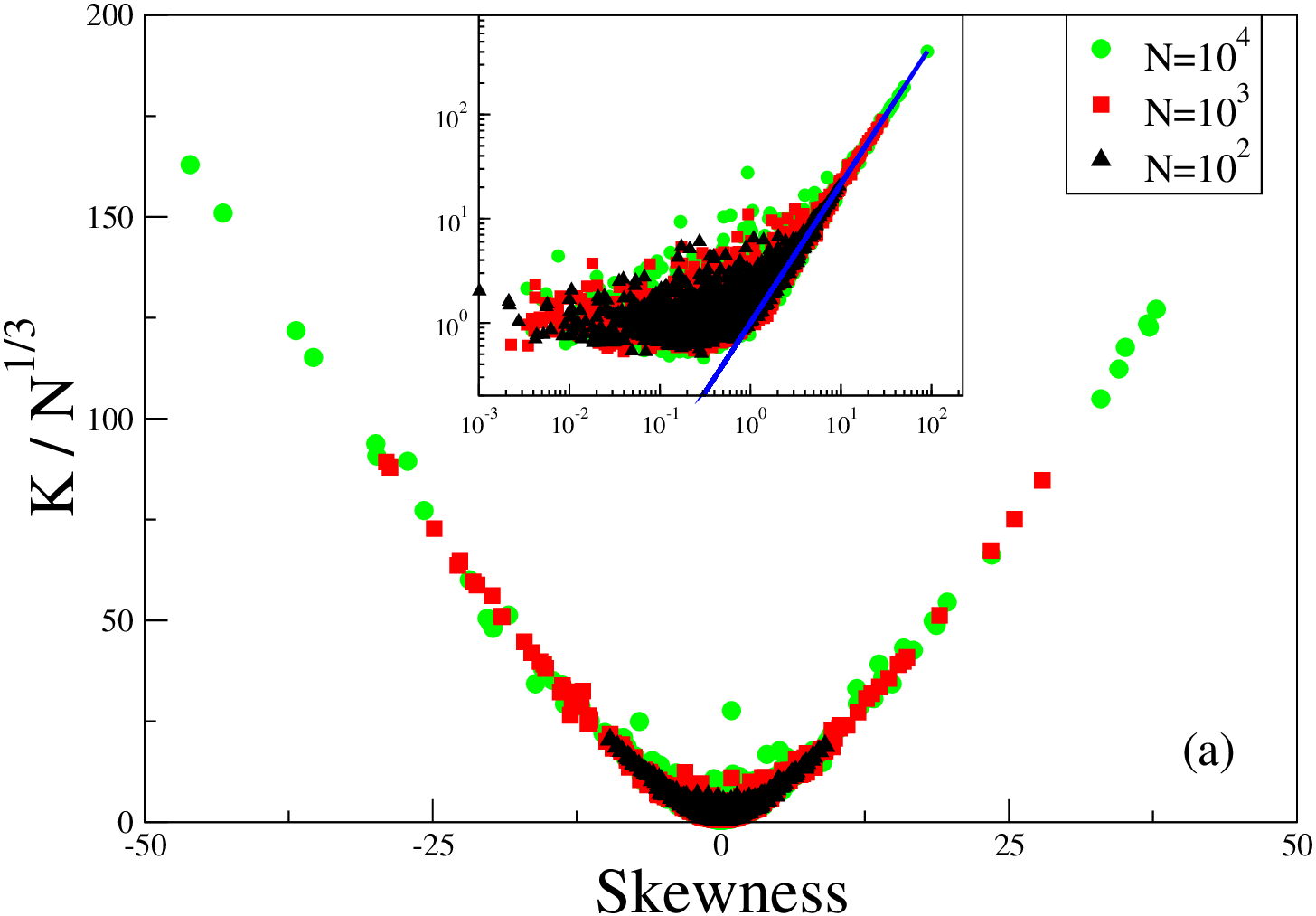} 
\includegraphics[width=8.1cm]{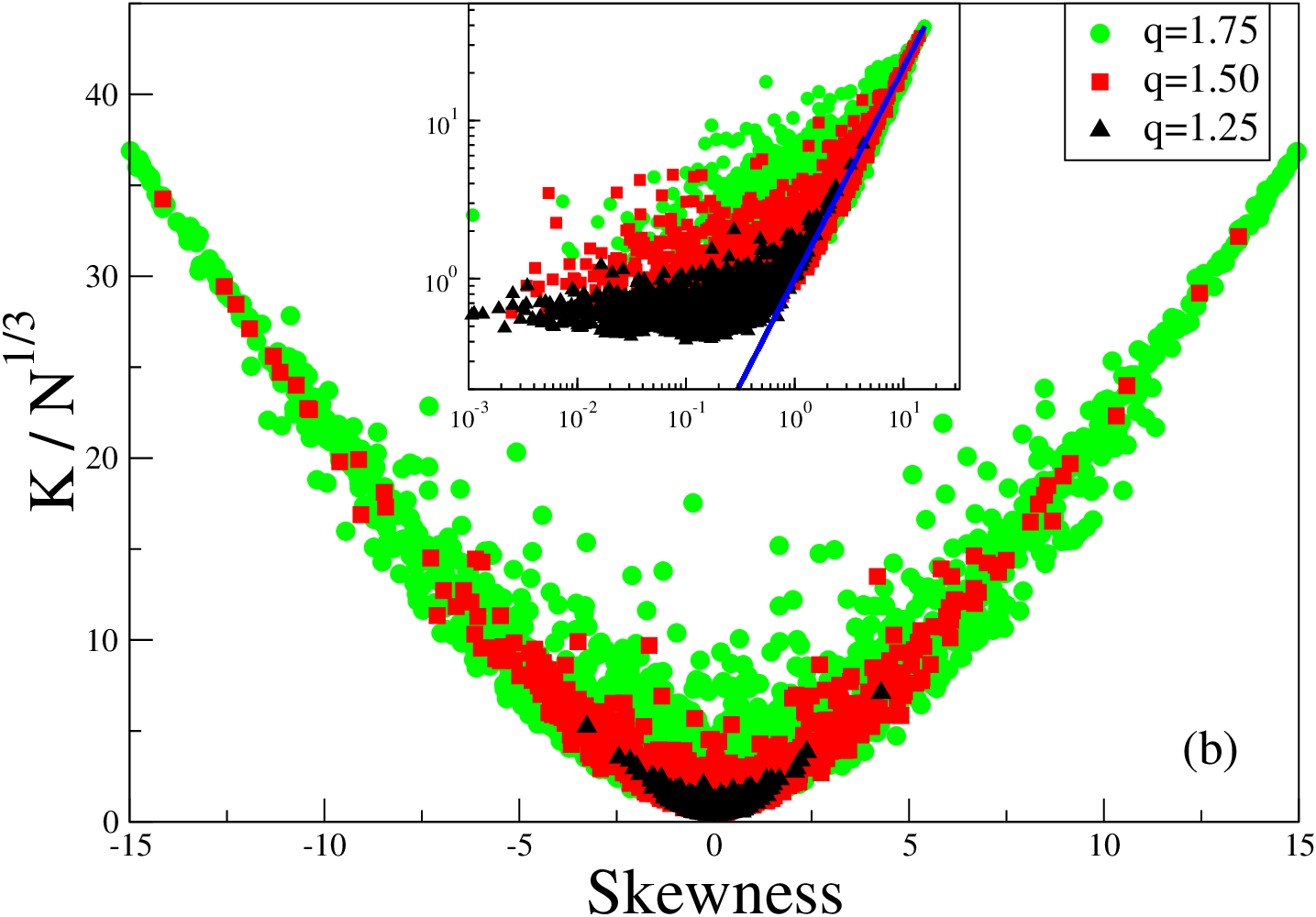}
\end{center}
\caption{(Color online) Kurtosis versus skewness for $q=1.5$ for different values of $N$ (a), and 
for different values of $q$ with $N=250$ (b). 
Three different curves have been collapsed onto each other by dividing the ordinate by $N^{1/3}$. 
In the Insets, we plot the same data in $\log$-$\log$ scale. The solid blue line is the power-law 
with exponent $4/3$.}
\label{fig:fig3}
\end{figure*}

The dashed black line in Fig.~\ref{fig:fig2} represents the power-law relation with exponent 2 (Eq.~(\ref{parabol})) 
and it corresponds to the lower bound of the points in $S-K$ plane. 
As shown in Fig.~\ref{fig:fig2}, the solid blue line (Eq.~(\ref{power})) matches quite well with the points away 
from this region. Thus our synthetic data for $q=1.5$ mimics exactly the bahavior 
of the economics data given in \cite{Cristelli} if the same number of data points is used.

Now we are at the position to test the behaviour for different values of $N$ with the same $q$ value and also for 
different values of $q$ with the same $N$ in order to understand whether and how $N$ and $q$ affect 
the behaviour of the system in the $S-K$ plane. Firstly, as shown in Fig.~\ref{fig:fig3}a, we plot kurtosis versus 
skewness for different $N$ values with the same $q$ value ($q=1.5$) and it is evident that, as $N$ increases, 
kurtosis reaches higher values following the power-law regime with exponent $4/3$. 
Then, as shown in Fig.~\ref{fig:fig3}b, 
we plot the same graph, this time, for different $q$ values fixing $N$ as $N=250$, where one can easily see that the 
same power-law behavior is reached for higher kurtosis values as $q$ increases. Therefore, increasing the 
values of $N$ and $q$ gives the same result although the mechanisms which cause this result are different. 
If $q$ is increased, this causes the distribution to be much more 
long-tailed and it is possible to find the extreme events more frequently and this of course gives the same result 
as increasing $N$, since, if you increase $N$ for a given $q$, the possibility of finding extreme events will 
also increase regardless of the value of $q$. As seen in Fig.~\ref{fig:fig2} and Fig.~\ref{fig:fig3}a, although all 
data in each window is selected from the same distribution (each window is a part of the same data set), some windows 
which have frequently more data from the tails have larger kurtosis values. This result does not necessarily indicate 
that the distribution with larger kurtosis value is more long tailed and therefore much more far away from the Gaussian. 
It only indicates that the contribution of tails in the summation for corresponding data set (window) is more than 
the other window. Moreover, let us suppose two distributions with different $q$ values. If one wants to compare 
these distributions, exactly opposite results for different $N$ values can be deduced. Of course, the distribution 
with larger $q$ exhibits more long-tailed distribution. Therefore, normally one expects that the 
distribution with larger $q$ value has a larger kurtosis. 
But, since extreme events are randomly distributed in the whole data set, 
there is no guarantee for a good representation of the tails for any given window 
in this considered data set. 
Sometimes, especially for small $N$ values, almost all data might come from the central part of the distribution. 
In Fig.~\ref{fig:fig3}b, each 
point represents a kurtosis and a skewness value of one window which has $N=250$. As seen in this figure, one can 
find several points in the power-law region with exponent $4/3$ of the distribution for $q=1.5$, whose kurtosis 
values are larger than all points of the power-law region with exponent 2 of the distribution for $q=1.75$. 
As a result, if one has these two different data sets which correspond to aforementioned windows, his/her conclusion 
would be that the distribution with $q=1.75$ is closer to the Gaussian than the distribution with $q=1.5$. 
But clearly this will be an incorrect conclusion. It only indicates that the distributions are not sampled 
sufficiently well in order for characterizing them correctly. This problem can only be overcome by using 
sufficiently large $N$. It is also worth mentioning that in Fig.~\ref{fig:fig3} the minimum of the power-law 
region with exponent 2 slowly increases as $N$ is increased for a given $q$ value or as the $q$ value of the 
distribution is increased for a given $N$ value.

\begin{figure*}[!th]
\begin{center}
\includegraphics[width=7.80cm]{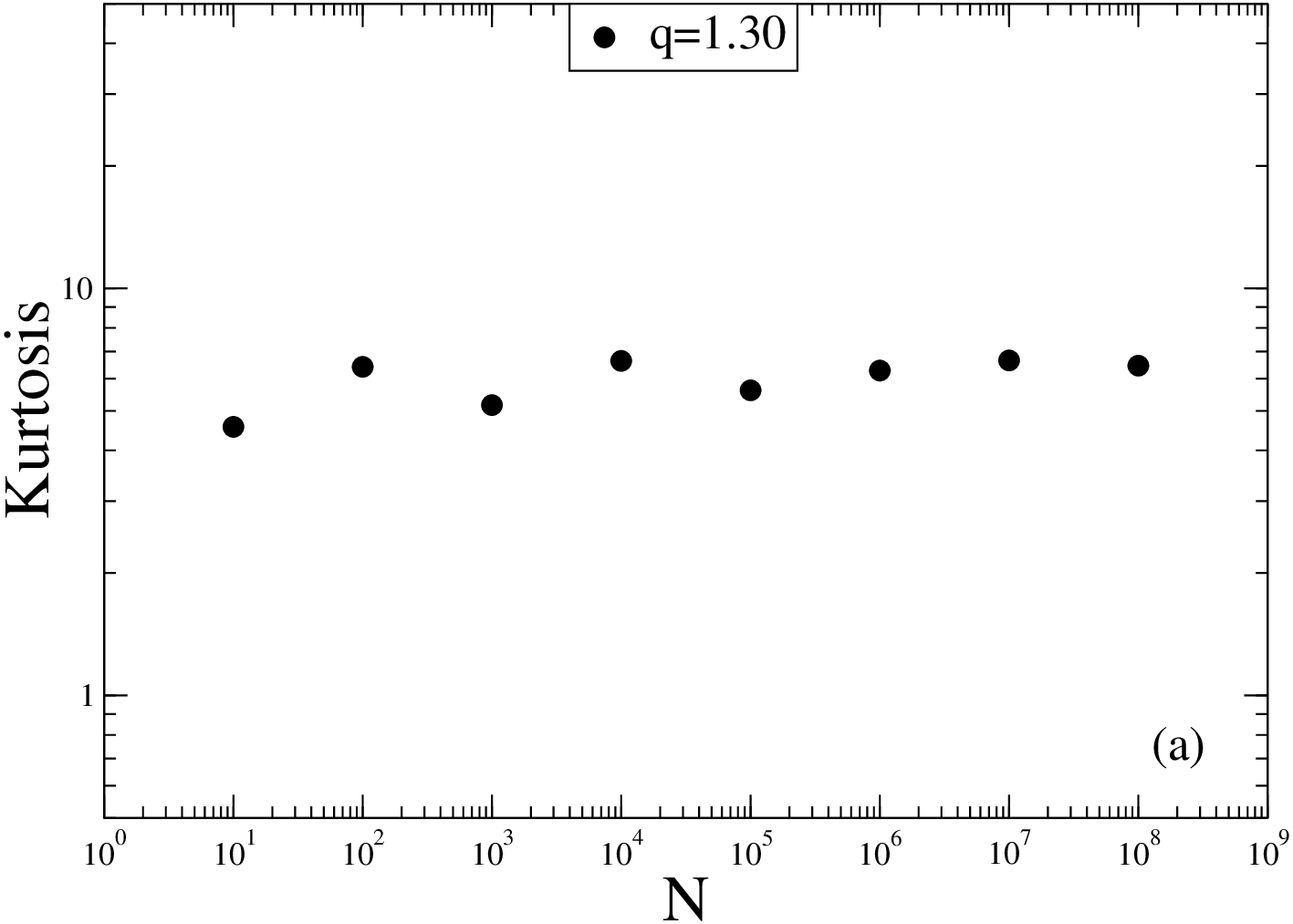} %
\includegraphics[width=7.80cm]{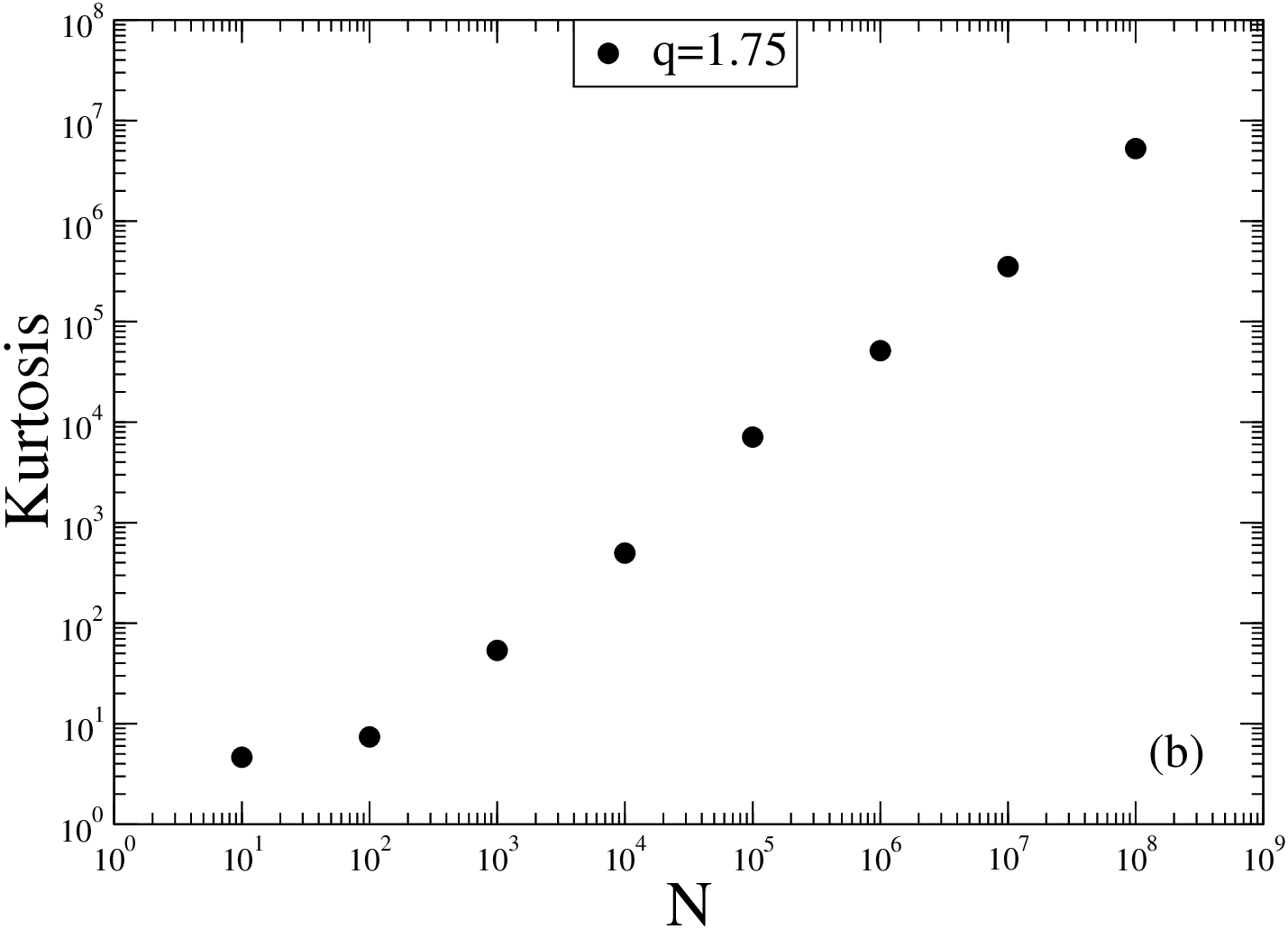} 
\end{center}
\caption{(Color online) Kurtosis as a function of $N$ for $q=1.3$ (a) and for $q=1.75$ (b).}
\label{fig:fig4}
\end{figure*}

The next step at this point must be to check whether the kurtosis approaches a finite value for any finite $N$. 
Therefore, we plot kurtosis as a function of $N$ for different $q$-Gaussians. The results of two representative 
cases are given in Fig.~\ref{fig:fig4}. It is clearly seen that the kurtosis approaches a fixed value as $N$ is 
moderately large for $q=1.3$ case, whereas the kurtosis does not saturate even for very large $N$ values like 
$10^8$ for $q=1.75$ case. We systematically checked several $q$-Gaussian distributions with $q\in[1,1.5]$ and observed 
that the kurtosis steadily increases with increasing $N$ and then it achieves a constant value. 
This value of $N$ at which $K$ saturates and also the value to which $K$ saturates increase with increasing 
$q$ as the distribution gets more and more distant from the Gaussian.

At this point, we conjecture that the kurtosis will always reach a constant value for sufficiently large $N$ if $q<5/3$, 
where the distribution has always a finite second moment. 
In fact, as seen in Fig.~\ref{fig:fig4}b for $q=1.75$ case, for larger $q$ values (namely, $5/3<q<3$), where 
the distribution exhibits infinite second moment, the $N$ values needed for constant kurtosis are 
postponed to very large values, most probably almost impossible to reach. 
After these findings, the plausible conclusion is that one can only compare two $q$-Gaussian distributions using the 
constant values of kurtosis if $1<q<5/3$. 
If the constant value of kurtosis is not obtained for a given distribution with $q\in[1,5/3]$ or if the 
distribution is from the region $q\in[5/3,3]$, then this comparison cannot be done.

\begin{figure*}[!th]
\begin{center}
\includegraphics[width=7.80cm]{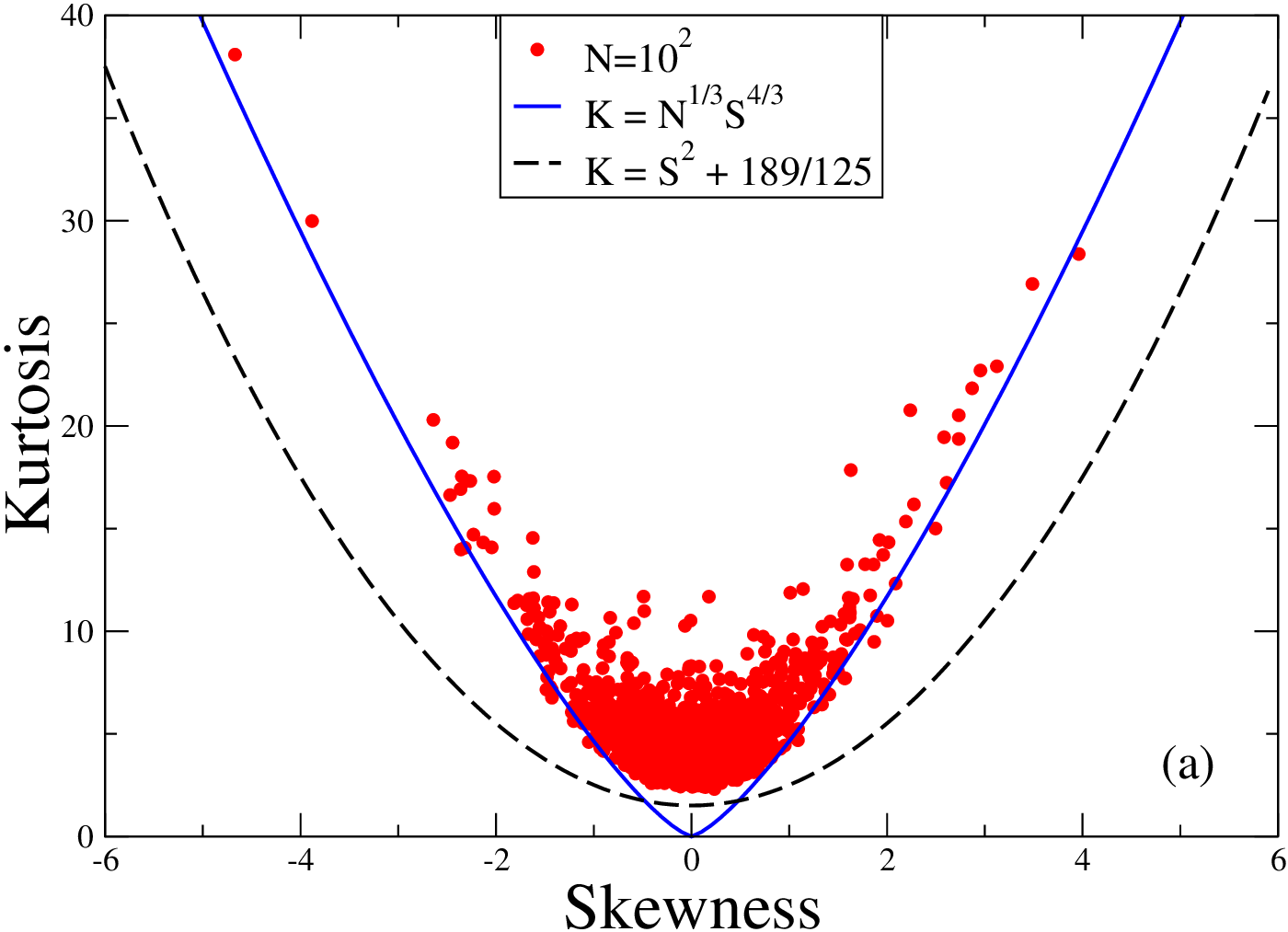} %
\includegraphics[width=7.80cm]{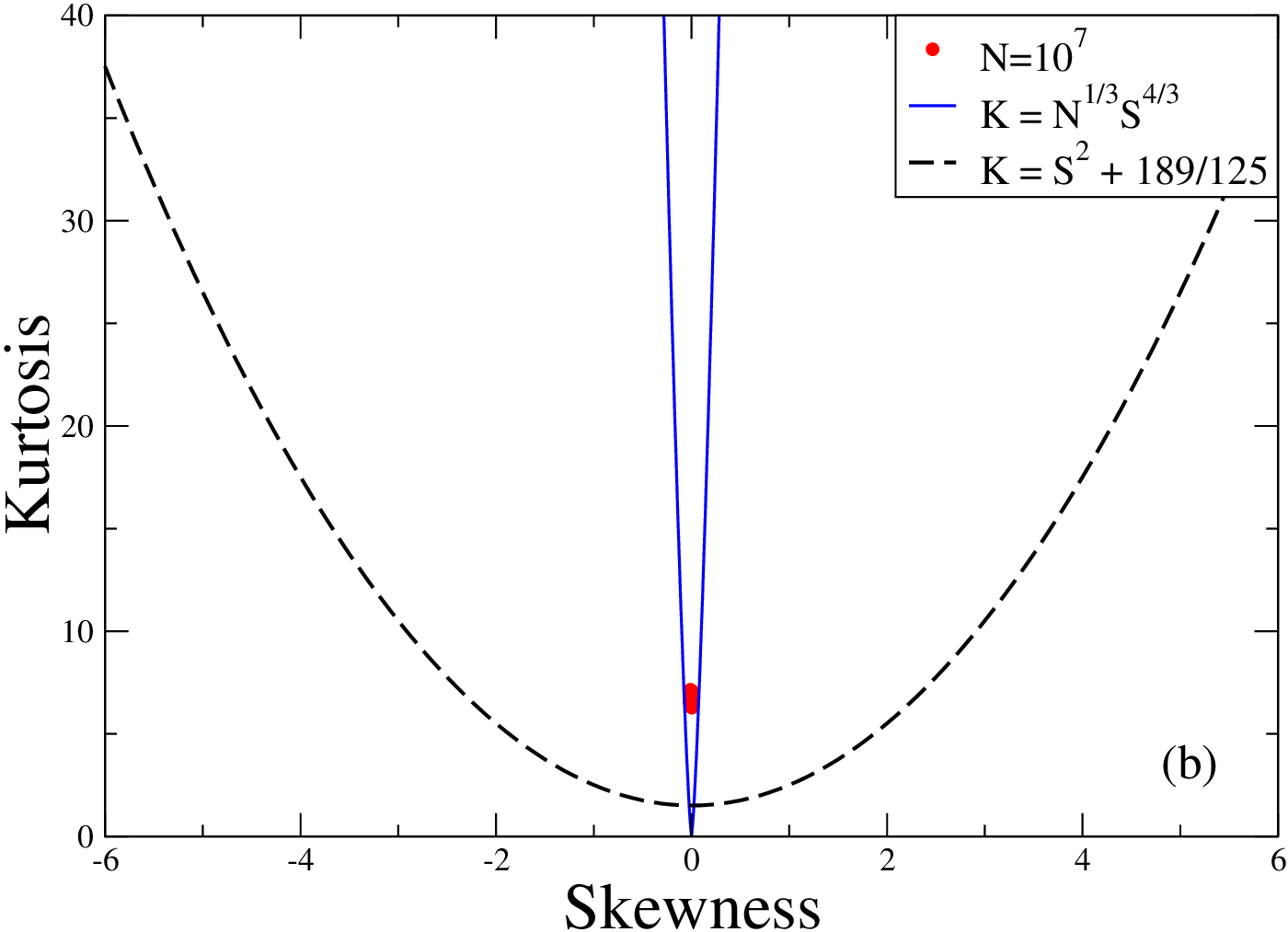} 
\end{center}
\caption{(Color online) Kurtosis versus skewness for $N=10^{2}$ (a) and for $N=10^{7}$ (b). 
Blue solid line and black dashed line refer to the power-law relations with exponent $4/3$ (Eq.~(\ref{power})) and 
with exponent 2 (Eq.~(\ref{parabol})) respectively.}
\label{fig:fig5}
\end{figure*}

Another question which needs to be answered is whether these two different power-law regimes are still valid for 
large $N$ for a $q$-Gaussian with $1<q<5/3$. To clarify this situation we plot again kurtosis as a function of 
skewness for $N=10^2$ and $N=10^7$ in Fig.~\ref{fig:fig5}. The $q$ is chosen as $1.3$ in order to reach the 
saturation of the kurtosis at an attainable $N$. As seen from Fig.~\ref{fig:fig5}a, for small $N$ case, 
two regimes are still valid since the kurtosis has not yet saturated to a constant value. 
On the other hand, in Fig.~\ref{fig:fig5}b, for large $N$ case, all points in the $S-K$ plane seem to converge 
to a single point and both regimes seem to disappear. 
One might only argue that the power-law relation with exponent 2 is valid as a lower bound.

\section{Levy as non-Gaussian distributions}

\subsection{Generating Levy Distributions}

Symmetric $\alpha$-stable Levy distributions, $L_{\alpha}(x)$, can be defined through its 
Fourier transform \cite{levy}
 
\begin{equation}
{\hat L}_{\alpha}(k) = \exp\left(-a |k|^{\alpha}\right) 
\end{equation}
where $\alpha$ is the Levy parameter in the interval $0<\alpha\le 2$. The case $\alpha=2$ 
corresponds to the standard Gaussian. 
Performing the inverse Fourier transform, one can evaluate the Levy distributions as 

\begin{equation}
L_{\alpha}(x) = \frac{1}{2\pi} \int_{-\infty}^{\infty} e^{-ikx} {\hat L}_{\alpha}(k) dk . 
\end{equation}
These distributions are known to exhibit infinite second moments for $0<\alpha <2$.

\begin{figure*}[!th]
\begin{center}
\includegraphics[width=10cm]{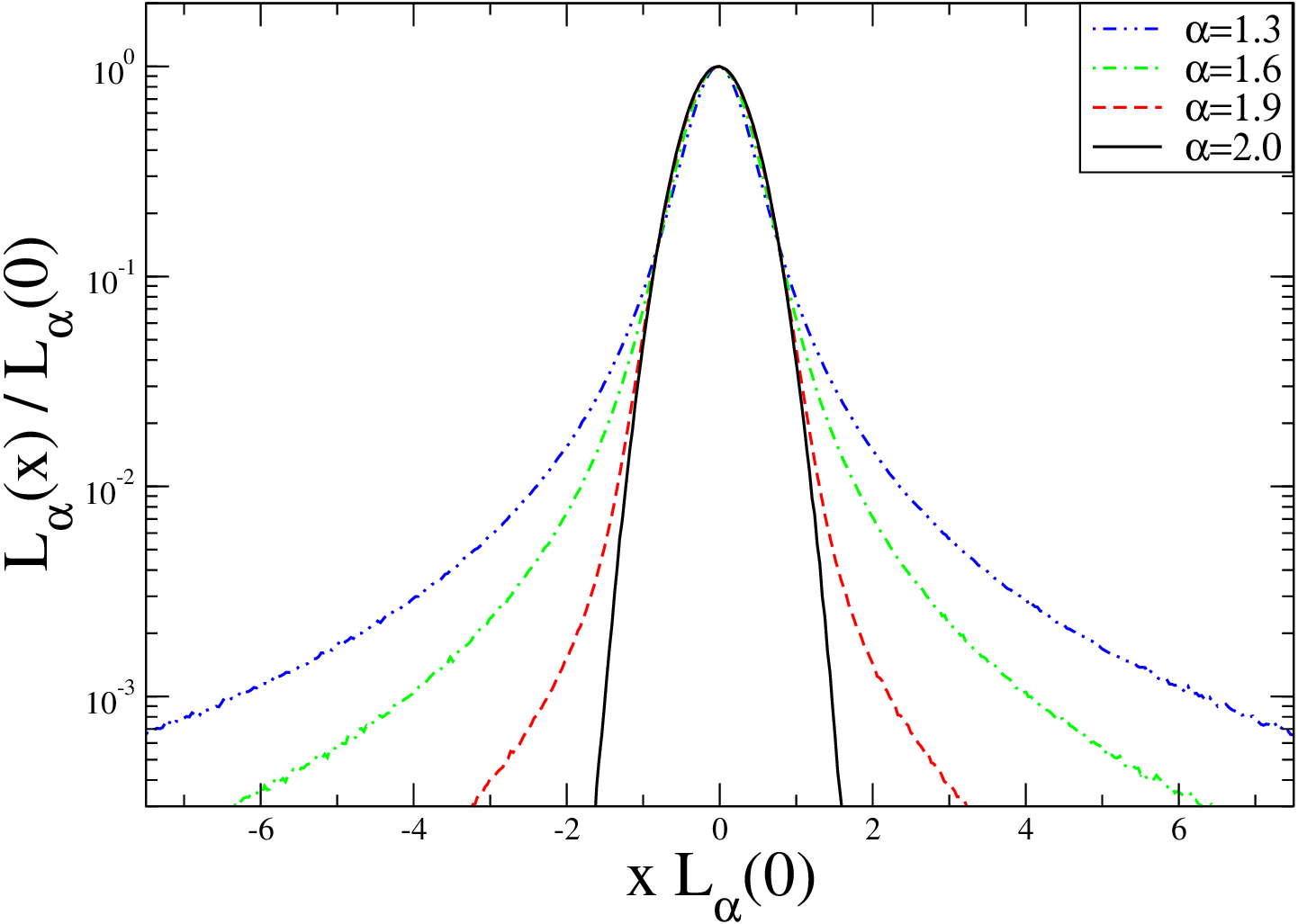} 
\end{center}
\caption{(Color online) Levy distributions for representative $\alpha$ values.  
Gaussian distribution is also given as a special case of $\alpha$ value ($\alpha=2$).}
\label{fig:fig6}
\end{figure*}

In order to generate $\alpha$-stable Levy distributions we use the MATLAB functions STBL given 
by M. Veillette \cite{veil}. For some representative $\alpha$ values, Levy distributions are 
plotted in Fig.~\ref{fig:fig6}. Using this function one can generate arbitrarily large number 
of $\alpha$-stable distributions which can be used in the same way that we have already done for 
$q$-Gaussians.

\subsection{Skewness and Kurtosis}

\begin{figure*}[!th]
\begin{center}
\includegraphics[width=10cm]{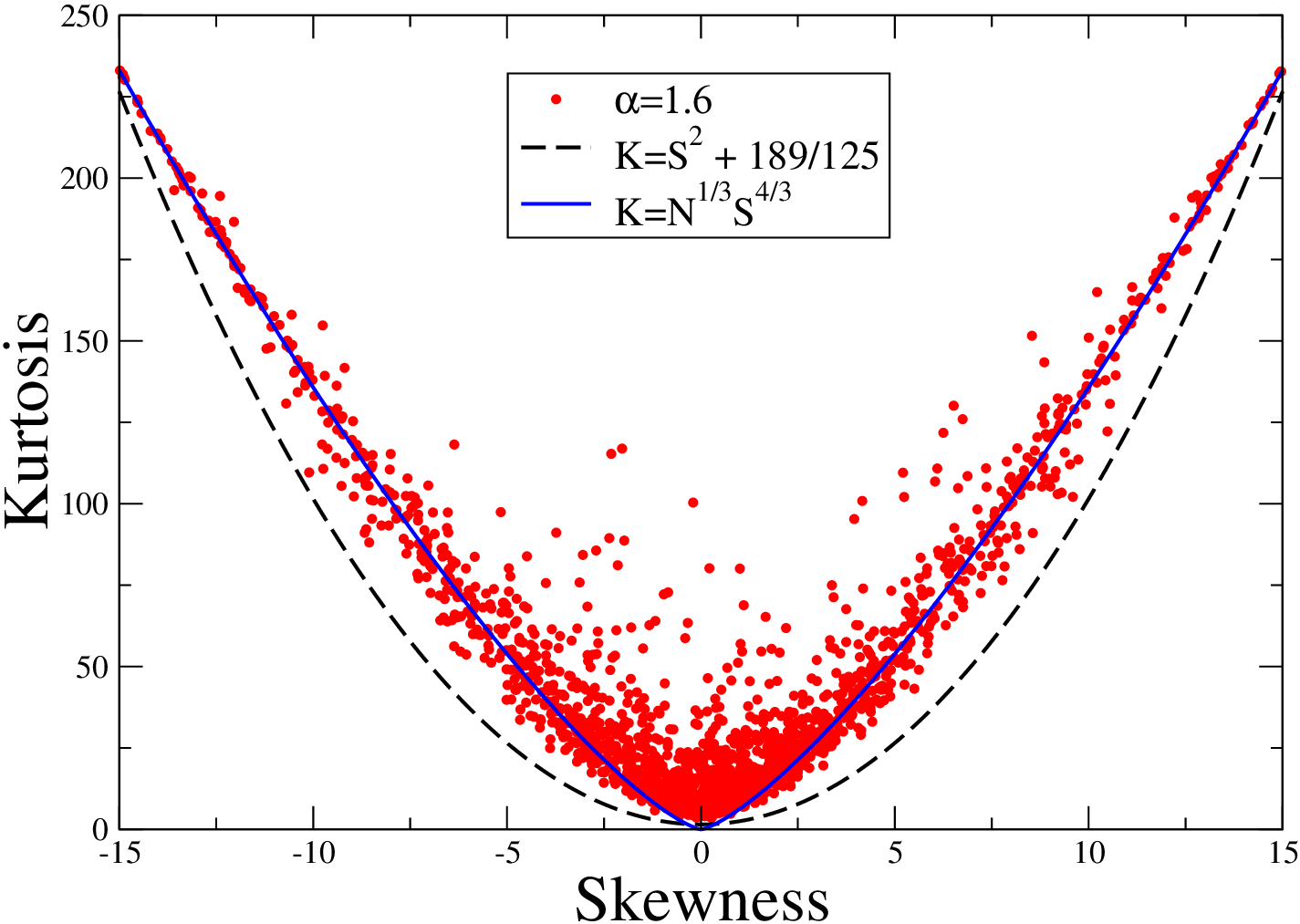}  
\end{center}
\caption{(Color online) Kurtosis versus skewness for $\alpha$-stable Levy distribution with $\alpha=1.6$. 
The number of points is $2000$. 
Each of them refers to a sample window and each window has $N=250$ number of data. }
\label{fig:fig7}
\end{figure*}

For $\alpha$-stable Levy distributions, we firstly analyze the kurtosis versus skewness behavior for 
small $N$ (namely, $N=250$) in Fig.~\ref{fig:fig7}. The similar tendency observed also for $q$-Gaussians 
is clearly evident. The power-law region with exponent 2 is followed only for a very short interval 
near $S=0$, whereas the other power-law region with exponent $4/3$ is realized for the values away 
from $S=0$.

\begin{figure*}[!th]
\begin{center}
\includegraphics[width=7.80cm]{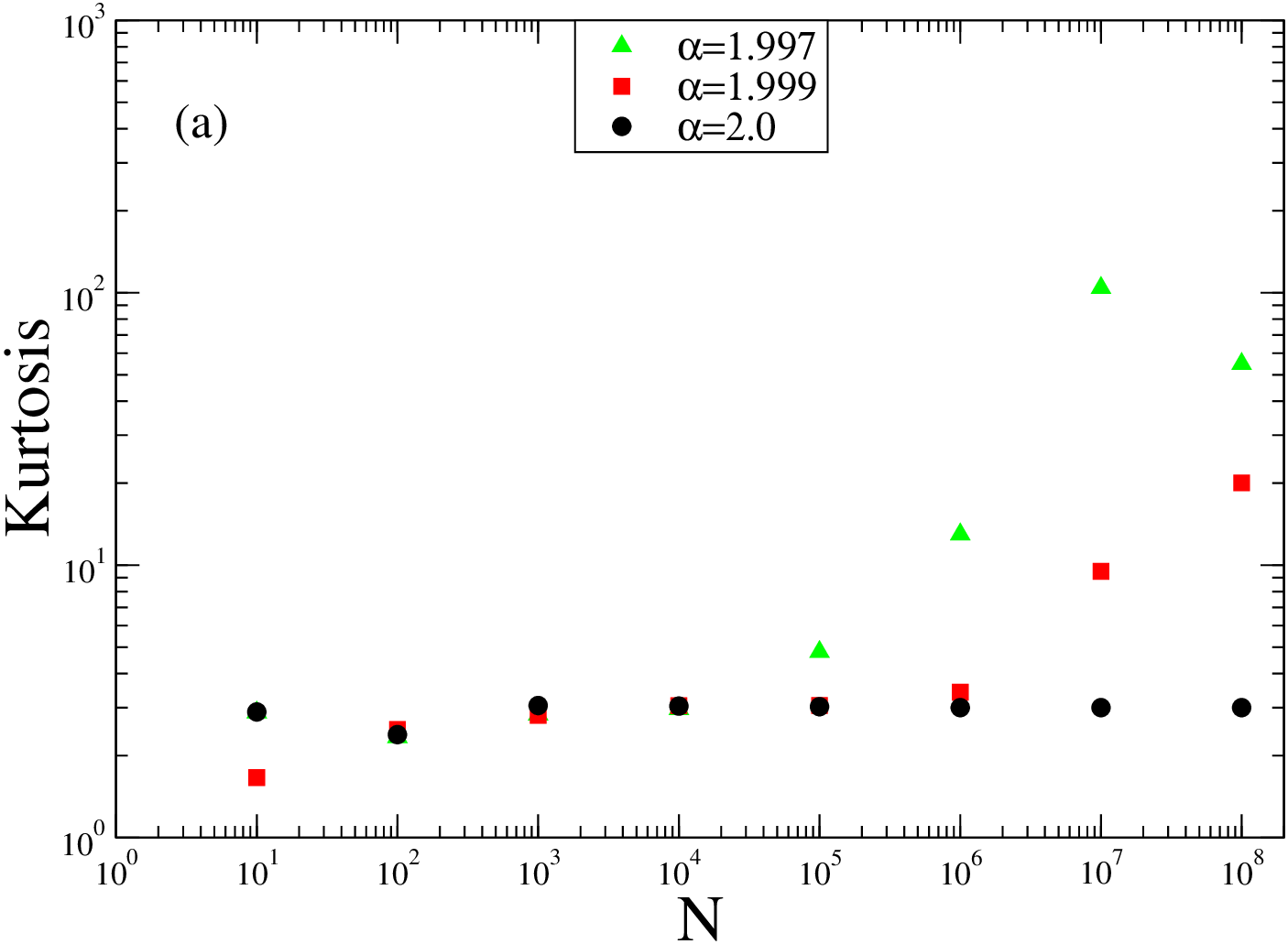} %
\includegraphics[width=7.80cm]{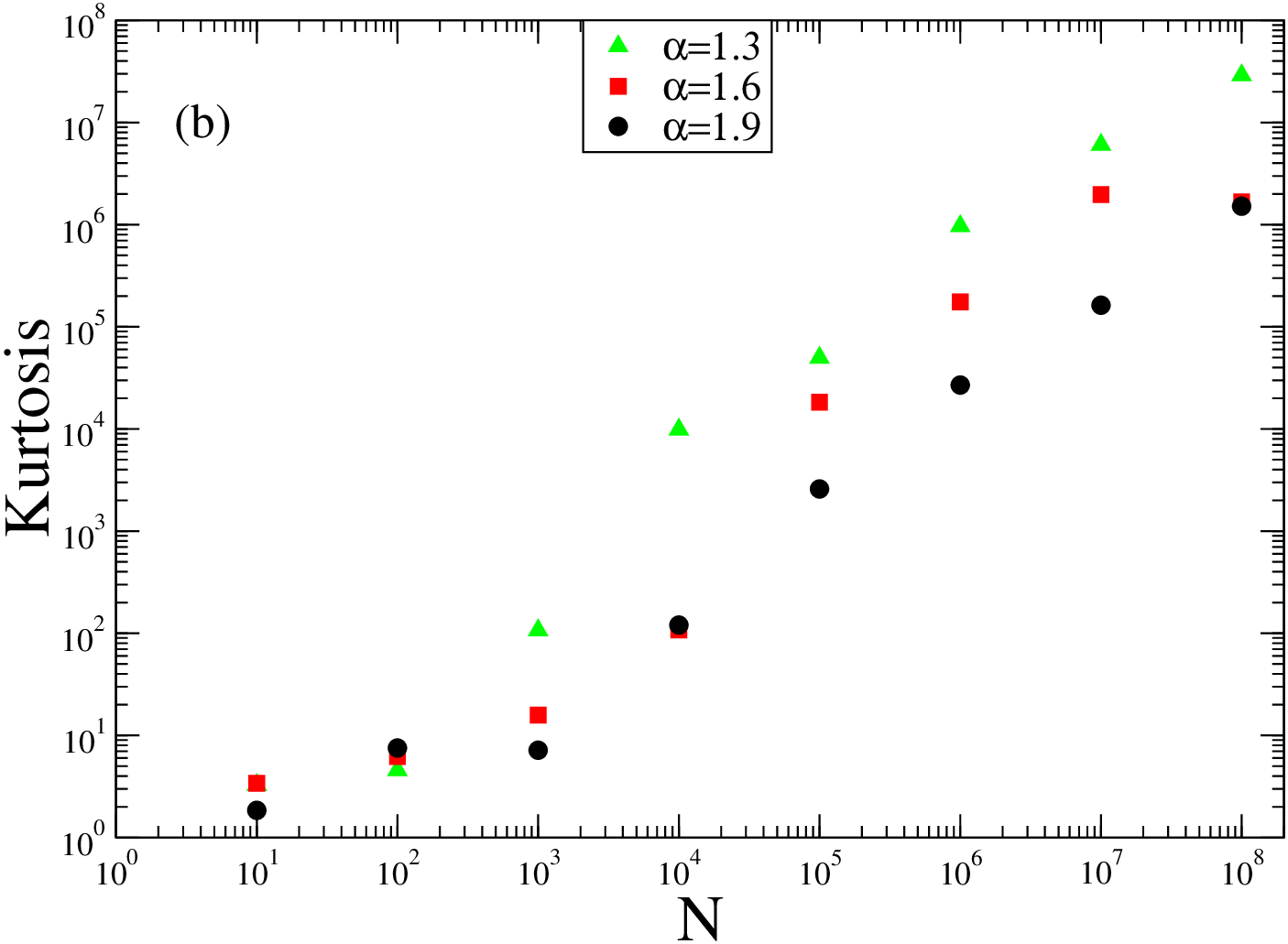} 
\end{center}
\caption{(Color online)  Kurtosis as a function of $N$ for $\alpha=2$, $\alpha=1.999$, $\alpha=1.997$ (a) 
and for $\alpha=1.3$, $\alpha=1.6$ and $\alpha=1.9$ (b).}
\label{fig:fig8}
\end{figure*}

Finally, we should check the behavior of kurtosis as a function of $N$. From what we understand in the 
$q$-Gaussian discussion, we must expect that, for all Levy distributions with $0<\alpha<2$, kurtosis values 
will never reach a constant value since the second moments of these distributions are always infinite. 
In order to see this better, we present in Fig.~\ref{fig:fig8} two representative cases for one of which 
we plot the results of distributions very close to Gaussian (see Fig.~\ref{fig:fig8}a) and the other is 
with the results of distributions far away from Gaussian (see Fig.~\ref{fig:fig8}b). 
For both cases, it is evident that the only constant kurtosis value ($K=3$) happens to be seen for $\alpha=2$ 
which is the Gaussian. It seems that, even if the distribution is very close to Gaussian, kurtosis will 
never attain a constant value for a finite $N$.

\section{CONCLUSION}

We check the relation between skewness and kurtosis in complex dynamics proposed in \cite{Cristelli} using synthetic 
large data sets for non-Gaussian distributions, namely, $q$-Gaussian and Levy distributions. 
Our results clearly show that the relation, proposed as being universal, happens to occur only due 
to insufficient number of data points used in the analysis if the second moment of the distribution is finite. 
This is just because the original distribution is not sampled sufficiently well for small $N$ values. 
We also verify that, as $N$ increases, the kurtosis value also increases up to some value of 
$N$ after which it remains constant. This $N$ value is postponed to larger values as the non-Gaussian 
distribution under consideration becomes more and more distant from Gaussian. If the second moment of the distribution 
is infinite, then this $N$ value seems to diverge. 
In the light of these findings, we conclude that using kurtosis to compare two different distributions 
(data sets) might lead incorrect results if the data sets in hand are not sufficiently large for distributions 
with finite second moment. 
To do this comparison correctly, one needs to be sure about the length of the data set which allows kurtosis 
to become a constant. 
If this is not guaranteed, then one might easily misinterpret a longer tailed distribution to be much closer to 
Gaussian compared to a shorter tailed one. 
On the other hand, if the distribution has infinite second moment like $q$-Gaussians with 
$5/3<q<3$ and all Levy distributions with $0<\alpha<2$, then it is completely irrelevant to compare these 
distributions using kurtosis.

Pearson has introduced kurtosis of a given distribution as a measure of deviation from Gaussian while he was trying 
to understand whether a distribution was Gaussian or not. Most probably it will be useful to replace the reference 
distribution (Gaussian) by a different one (non-Gaussian), which is much closer to the distribution under 
consideration. 
We are planning to discuss this issue elsewhere in the near future.

\section*{Acknowledgment}
We are indebted to Ayse Erzan for very fruitful discussions and useful remarks. 
This work has been supported by TUBITAK (Turkish Agency) under the Research Project number 112T083. 
U.T. is a member of the Science Academy, Istanbul, Turkey.

\end{document}